\documentclass[twoside,twocolumn,9pt]{article}
\usepackage{extsizes}
\usepackage[super,sort&compress,comma]{natbib} 
\usepackage[version=3]{mhchem}
\usepackage[left=1.5cm, right=1.5cm, top=1.785cm, bottom=2.0cm]{geometry}
\usepackage{balance}
\usepackage{mathptmx}
\usepackage{sectsty}
\usepackage{graphicx} 
\usepackage{lastpage}
\usepackage[format=plain,justification=justified,singlelinecheck=false,font={stretch=1.125,small,sf},labelfont=bf,labelsep=space]{caption}
\usepackage{float}
\usepackage{fancyhdr}
\usepackage{fnpos}
\usepackage[english]{babel}
\addto{\captionsenglish}{%
  
}
\usepackage{array}
\usepackage{droidsans}
\usepackage{charter}
\usepackage[T1]{fontenc}
\usepackage[usenames,dvipsnames]{xcolor}
\usepackage{setspace}
\usepackage[compact]{titlesec}
\usepackage{hyperref}
\usepackage{gensymb}
\usepackage[pagewise, modulo]{lineno}

\usepackage{epstopdf}

\definecolor{cream}{RGB}{222,217,201}

\begin{document}

\pagestyle{fancy}
\thispagestyle{plain}
\fancypagestyle{plain}{
\renewcommand{\headrulewidth}{0pt}
}

\makeFNbottom
\makeatletter
\renewcommand\LARGE{\@setfontsize\LARGE{15pt}{17}}
\renewcommand\Large{\@setfontsize\Large{12pt}{14}}
\renewcommand\large{\@setfontsize\large{10pt}{12}}
\renewcommand\footnotesize{\@setfontsize\footnotesize{7pt}{10}}
\makeatother

\renewcommand{\thefootnote}{\fnsymbol{footnote}}
\renewcommand\footnoterule{\vspace*{1pt}%
\color{cream}\hrule width 3.5in height 0.4pt \color{black}\vspace*{5pt}} 
\setcounter{secnumdepth}{5}

\makeatletter 
\renewcommand\@biblabel[1]{#1}            
\renewcommand\@makefntext[1]%
{\noindent\makebox[0pt][r]{\@thefnmark\,}#1}
\makeatother 
\renewcommand{\figurename}{\small{Fig.}~}
\sectionfont{\sffamily\Large}
\subsectionfont{\normalsize}
\subsubsectionfont{\bf}
\setstretch{1.125} 
\setlength{\skip\footins}{0.8cm}
\setlength{\footnotesep}{0.25cm}
\setlength{\jot}{10pt}
\titlespacing*{\section}{0pt}{4pt}{4pt}
\titlespacing*{\subsection}{0pt}{15pt}{1pt}

\fancyfoot{}
\fancyfoot[LO,RE]{\vspace{-7.1pt}\includegraphics[height=9pt]{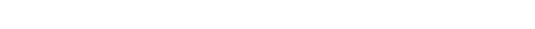}}
\fancyfoot[CO]{\vspace{-7.1pt}\hspace{11.9cm}\includegraphics{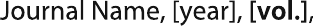}}
\fancyfoot[CE]{\vspace{-7.2pt}\hspace{-13.2cm}\includegraphics{head_foot/RF}}
\fancyfoot[RO]{\footnotesize{\sffamily{1--\pageref{LastPage} ~\textbar  \hspace{2pt}\thepage}}}
\fancyfoot[LE]{\footnotesize{\sffamily{\thepage~\textbar\hspace{4.65cm} 1--\pageref{LastPage}}}}
\fancyhead{}
\renewcommand{\headrulewidth}{0pt} 
\renewcommand{\footrulewidth}{0pt}
\setlength{\arrayrulewidth}{1pt}
\setlength{\columnsep}{6.5mm}
\setlength\bibsep{1pt}

\makeatletter 
\newlength{\figrulesep} 
\setlength{\figrulesep}{0.5\textfloatsep} 

\newcommand{\topfigrule}{\vspace*{-1pt}%
\noindent{\color{cream}\rule[-\figrulesep]{\columnwidth}{1.5pt}} }

\newcommand{\botfigrule}{\vspace*{-2pt}%
\noindent{\color{cream}\rule[\figrulesep]{\columnwidth}{1.5pt}} }

\newcommand{\dblfigrule}{\vspace*{-1pt}%
\noindent{\color{cream}\rule[-\figrulesep]{\textwidth}{1.5pt}} }

\makeatother

\twocolumn[
  \begin{@twocolumnfalse}
{\includegraphics[height=30pt]{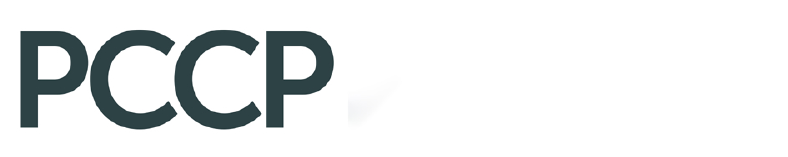}\hfill\raisebox{0pt}[0pt][0pt]{\includegraphics[height=55pt]{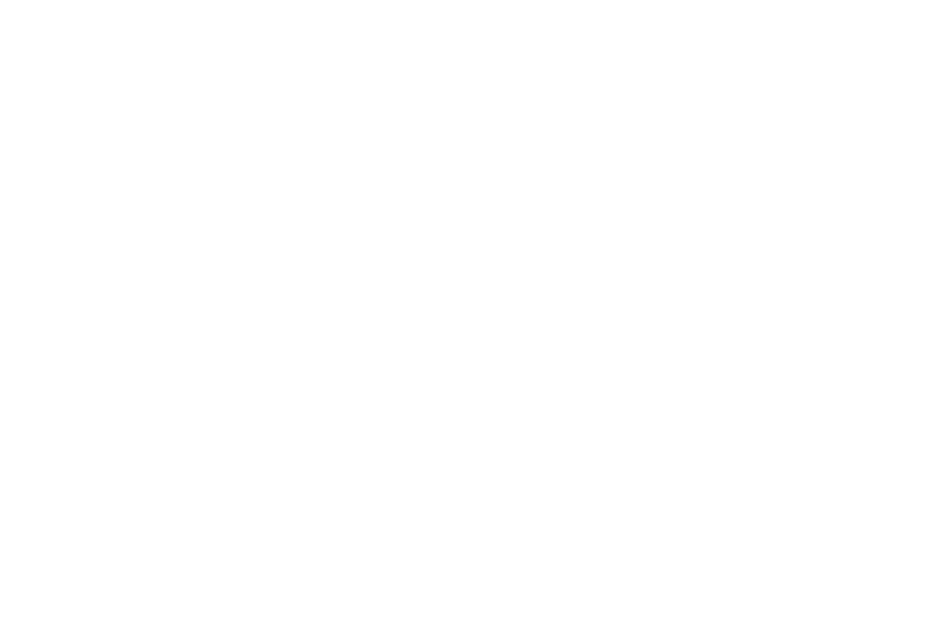}}\\[1ex]
\includegraphics[width=18.5cm]{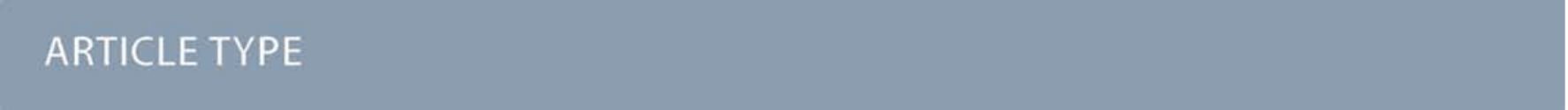}}\par
\vspace{1em}
\sffamily
\begin{tabular}{m{4.5cm} p{13.5cm} }

\includegraphics{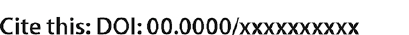} & \noindent\LARGE{\textbf{IR photofragmentation of the Phenyl Cation: Spectroscopy and Fragmentation Pathways$^\dag$
}} \\
\vspace{0.3cm} & \vspace{0.3cm} \\

 & \noindent\large{Sandra D. Wiersma,\textit{$^{a,b,e\ddag}$} Alessandra Candian,\textit{$^{a\ddag}$} Joost M. Bakker,\textit{$^{b}$} Giel Berden,\textit{$^{b}$} John R. Eyler,\textit{$^{d}$}  Jos Oomens,\textit{$^{b}$}, Alexander G. G. M. Tielens,\textit{$^{c}$} Annemieke Petrignani$^{\ast}$\textit{$^{a,c}$}} \\

\includegraphics{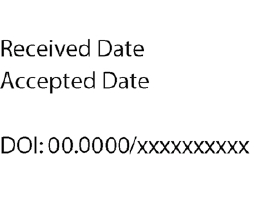} & \noindent\normalsize{We present the gas-phase infrared spectra of the phenyl cation, phenylium, in its perprotio \ce{(C6H5+)} and perdeutero \ce{(C6D5+)} forms, in the 260--1925 cm$^{-1}$ (5.2--38 $\mu$m) spectral range, and investigate the observed photofragmentation. The spectral and fragmentation data were obtained using Infrared Multiple Photon Dissociation (IRMPD) spectroscopy within a Fourier Transform Ion Cyclotron Resonance Mass Spectrometer (FTICR MS) located inside the cavity of the free electron laser FELICE (Free Electron Laser for Intra-Cavity Experiments). The \ce{^1A_1} singlet nature of the phenylium ion is ascertained by comparison of the observed IR spectrum with DFT calculations, using both harmonic and anharmonic frequency calculations. To investigate the observed loss of predominantly [2C,nH] (n$=2-4$) fragments, we explored the potential energy surface (PES) to unravel possible isomerization and fragmentation reaction pathways. The lowest energy pathways toward fragmentation include direct H elimination, and a combination of facile ring-opening mechanisms ($\leq2.4$ eV), followed by elimination of H or \ce{CCH2}. Energetically, all H-loss channels found are more easily accessible than \ce{CCH2}-loss. Calculations of the vibrational density of states for the various intermediates show that at high internal energies, ring opening is the thermodynamically the most advantageaous, eliminating direct H-loss as a competing process. The observed loss of primarily [2C,2H] can be explained through entropy calculations that show favored loss of [2C,2H] at higher internal energies.
} \\

\end{tabular}

 \end{@twocolumnfalse} \vspace{0.6cm}

  ]

\renewcommand*\rmdefault{bch}\normalfont\upshape
\rmfamily
\section*{}
\vspace{-1cm}


\footnotetext{\textit{$^{a}$~Van `t Hoff Institute for Molecular Sciences, University of Amsterdam, PO Box 94157, 1090 GD, Amsterdam, The Netherlands  E-mail: a.petrignani@uva.nl}}
\footnotetext{\textit{$^{b}$~Radboud University, Institute for Molecules and Materials, FELIX Laboratory, Toernooiveld 7, 6525 ED Nijmegen, The Netherlands }}
\footnotetext{\textit{$^{c}$~Leiden Observatory, Leiden University, P.O. Box 9513, 2300 RA Leiden, The Netherlands}}
\footnotetext{\textit{$^{d}$~Department of Chemistry, University of Florida, P.O. Box 117200, Gainesville, FL 32611-7200, U.S.A.}}
\footnotetext{\textit{$^{e}$~Institut de Recherche en Astrophysique et Plan\'etologie,  9 avenue du Colonel Roche, BP 44346, 31028 Toulouse Cedex 4, France}}

\footnotetext{\dag~Electronic Supplementary Information (ESI) available: optimised structures of intermediates and transition states, density of state calculations for phenylium, details about anharmonic calculations.}

\footnotetext{\ddag~These authors contributed equally to this work}


\section{Introduction}
The phenyl cation, C$_6$H$_5^+$, also known as phenylium, is a benchmark system for several organic species such as the aryl and arene groups. As an abundant component in  hydrocarbon plasmas, it plays a key role in flames and combustion chemistry where ring growth according to the HACA mechanism (H-abstraction/acetylene-addition, C$_2$H$_2$) takes place.\cite{Frenklach1989,2000:pope,2000:richter,2001:lindstedt,2002:frenklach,2018:ruwe,2019:vitiello} Phenyl and other benzene derivatives are thereby considered to be a cornerstone for aromatic growth, forming Polycyclic Aromatic Hydrocarbons (PAHs).\citep{2016:yang} The release of arylic compounds such as  C$_6$H$_5^+$ into Earth's atmosphere has sparked interest in the reaction mechanisms in which they could be involved.\cite{2019:mondal}

Larger PAHs containing more than 40 C atoms are known to be present in space, and to play a key role in interstellar chemistry.\cite{1992:cherchneff,2008:tielens,2011:peeters,2012:soliman,2013:tielens,2017:mebel} They are observed via infrared emission in characteristic bands, called Aromatic Infrared Bands (AIBs). Although individual PAH species are still to be identified, benzene\cite{2001:cernicharo,2006:kraemer}, and benzonitrile\cite{2018:mcguire} have been firmly identified in space. Benzene has also been detected in the atmospheres of Titan and on giant planets.\cite{2001:bezard,2005:lebonnois} Phenylium is currently one of the proposed precursors in the interstellar synthesis of benzene.\cite{2002:woods,2016:Peverati}

\begin{figure}
	\centering
	\includegraphics[width=\linewidth]{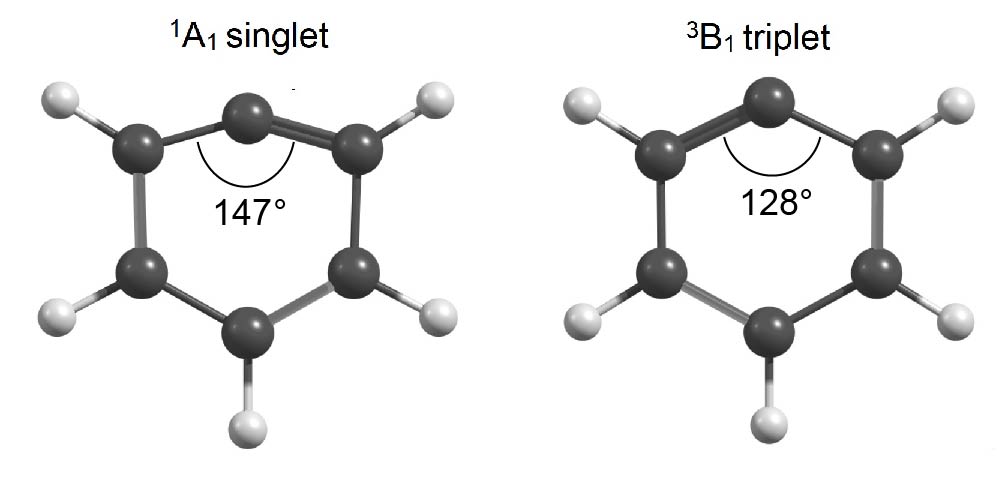}
	\caption{The geometries of the two different possible electronic states, showing how the singlet structure leads to a distorted hexagonal structure.  \label{fig:struct}}
\end{figure}
Phenylium's high reactivity stems from its electrophilicity, which is induced in the singlet state by its vacant, non-bonding $\sigma$ orbital\cite{1976:dill}. The empty $\sigma$ orbital induces sp hybridization of the C-atom, causing a substantial deformation of the hexagonal frame as is depicted in the left structure in Fig. \ref{fig:struct}. In its triplet state (right structure), the $\sigma$ orbital is singly occupied by an electron that has been removed from the $\pi$-system.\cite{1997:nicolaides} This restores the sp$^2$ hybridization of the carbon center and hence the hexagonal shape of the \ce{C6H5+} cation, but sacrifices the aromaticity of the system.\cite{1961:taft}
Whereas there exists a general agreement that the $^1$A$_1$ state is the ground state, the energy difference between the triplet and singlet states is somewhat uncertain, with calculations covering a range from 77 to 137 kJ mol$^{-1}$\cite{2016:Peverati,1997:nicolaides,1997:hrusak,1994:krauss,1976:dill}. Photoelectron spectroscopy of the phenyl radical did not shed any light on the singlet-triplet (S-T) gap.\cite{1987:butcher} Some experimental support of a singlet ground state was offered in 2000, when a matrix-isolated IR spectroscopy study of phenylium reported vibrational frequencies at 713 and 3110 cm$^{-1}$.\cite{2000:winkler} These two bands were found to be consistent with a scaled harmonic frequency prediction calculated with B3LYP/cc-pVDZ for the $^1$A$_1$ state. A decade later, a gas-phase, IR photofragmentation spectrum of argon-tagged phenylium was reported, presenting five bands in the 3 $\mu$m range characteristic for \ce{C-H} stretch vibrations.\cite{2010:patzer} The complementary calculations at the MP2/aug‐cc‐pVTZ and scaled MP2/6‐311++G(2df,2pd) levels of theory showed that these \ce{C-H} stretch vibrations shift, depending on the position of the Ar tag relative to the phenylium cation. In the $^1$A$_1$ state, the Ar atom donates electron density into the empty $\sigma$ orbital on the carbocation center and forms a $\sigma$ bond in the plane of the molecule. In the $^3$B$_1$ state, the Ar atom shares electron density with the $\pi$ system of the cation and localizes above the face of the ring. In 2012, an extensive experimental and theoretical investigation of several singly dehydrogenated PAH cations reported a linear correlation between the size of the PAH molecule and the size of the S-T gap, predicting triplet electronic ground states for all singly dehydrogenated PAHs and a singlet configuration for phenylium.\cite{2012:alvarogalue} Although experimental verification was presented for the larger aryl cations by applying IR Multiple Photon Dissociation (IRMPD) spectroscopy using the Free Electron Laser for Infrared Experiments (FELIX), no experimental spectrum of phenylium was reported. Phenylium proved too resistant against IRMPD, preventing the recording of the fingerprint IR spectrum of bare (\textit{i.e.,} untagged), gas-phase phenylium.

Experimental evidence pertaining to isomerization and fragmentation is likewise limited. Such information is of importance as a structural rearrangement of phenylium would affect reaction and dissociation pathways that are of importance to the reaction networks in flames and the interstellar medium. Recently, \citet{Shi:2016} theoretically explored the phenylium potential energy surface (PES), identifying 60 possible isomers in which they were able to connect 28 through transition states.
A laboratory study on the photodissociation of the (neutral) phenyl radical found H-loss to occur upon irradiation at 248 nm and both H and [2C,2H]-loss at 193 nm.\cite{Negru2010} The fragmentation of phenylium upon collisions with He and \ce{N2O} has also been reported to lead to [2C,2H]-loss.\cite{Ascenzi2007,Giordana2009} Furthermore, many studies on the larger PAH species have shown a trend for smaller and irregular PAHs to be more likely to lose [2C,2H] than \ce{H} or \ce{H2} upon photofragmentation.\cite{Jochims1994,Ekern1998,Zhen2014,Stockett2015,Jusko2018,West2018,West2019} It would thus be of great interest to find out if  phenylium follows this trend.

In this paper, we present the IRMPD spectrum of bare, isolated, gas-phase phenylium, in its perprotio \ce{(C6H5+)} and perdeutero \ce{(C6D5+)} forms in the 5--40 $\mu$m region. We use the FT-ICR mass spectrometer in the optical cavity of the IR Free Electron Laser for Intra-Cavity Experiments (FELICE),\cite{2016:petrignani} demonstrating that the higher fluences available there are able to efficiently excite IRMPD-resistant species.\cite{2010:bakker,Lapoutre2013,Wensink}$^{\dag}$ The observed fragmentation products are interpreted by investigating several isomerization and fragmentation pathways using DFT calculations.  

\section{Methods}

\subsection{Experiments}
The IR spectrum of phenylium was recorded in the Fourier Transform Ion Cyclotron Resonance Mass Spectrometer (FT-ICR MS) located inside the optical cavity of FELICE, one of the free-electron lasers of the FELIX Laboratory.\cite{2016:petrignani} A schematic of the apparatus can be found in \citet{2016:petrignani}; only details relevant to this study are given here.

Phenylium is generated in an electron impact ionizer. To form the dehydrogenated cation of benzene, bromobenzene (C$_6$H$_5$Br or \ce{C6D5Br} Merck/Sigma-Aldrich) is used as precursor and placed in a glass vial connected via a leak valve to the instrument, operated at source chamber pressures of several $10^{-7}$ mbar. The bromobenzene molecules are subsequently ionized with 40-eV electrons generating phenylium (C$_6$H$_5^+$ $m/z=77$, or \ce{C6D5+}, $m/z=82$). 

All ions are directed to a quadrupole mass selector, used in radio frequency (rf) only mode, to guide all ions. The continuously produced ions are then accumulated and collisionally cooled with room-temperature Ar gas at an ambient pressure of $10^{-2}$ mbar in a linear quadrupole ion trap with rectilinear rods segmented in three parts.\cite{2004:ouyang} The accumulated ions are extracted into a large electrostatic deflection quadrupole located inside the laser cavity. The ions are deflected by 90\degree~to be parallel to the laser beam into a 1-m-long, rf-guiding quadrupole, which leads to four ion trapping and detection cells positioned in the bore of the 7-T FT-ICR MS.\cite{2016:petrignani} The center of cell 1 is located at the focus of the FELICE laser beam, and cells 2, 3, and 4 are positioned progressively 100 mm away from the previous cell. The magnet can be moved along its axis so that one of the four ICR cells can be selected to be in the sweet spot of the magnet. Considering the Rayleigh range of 82 mm, the laser fluence is reduced by a factor of 2.3 for each next cell. The high photon densities provided in the laser focus were not required to achieve fragmentation. To limit power broadening effects, the experiments were therefore conducted in cell 4. For \ce{C6H5+} the ions were irradiated for 4.85 s at a repetition rate of 10 Hz amounting to 48 macropulses. After these experiments, development on the alignment optics made it possible to achieve the same results for \ce{C6D5+} with only two macropulses in 0.21 s at a repetition rate of 5 Hz.

During storage in cell 4, the desired mass, \textit{i.e.,} $m/z=77$ for \ce{C6H5+} or $m/z=82$ for \ce{C6D5+}, is isolated by ejecting all other masses present via a Stored Waveform Inverse Fourier Transform (SWIFT) pulse.\cite{Marshall1985} After irradiation of the ion cloud, a mass spectrum is recorded at each frequency step. The intensity for each fragment ion is recorded as function of FELICE frequency. The IRMPD spectrum is obtained by taking the ratio of the total fragment ion count and the total ion count, \textit{i.e.,} the sum of the parent and fragment ions, giving the yield $Y(\nu)$ 
\[Y(\nu)= \frac{\sum{I_{frag}(\nu)}}{I_{par}(\nu)+\sum{I_{frag}(\nu)}}\] 
This yield is then normalized to the macropulse energy that is obtained by monitoring the FELICE power. For this a fraction of the laser beam is coupled out through a 0.5 mm radius hole in the cavity end mirror and is directed onto a power meter (Coherent EPM1000). Wavelength calibration was performed by directing the IR beam to a grating spectrometer (Princeton Instruments SpectraPro). The spectral bandwidth is near transform-limited and is at $\sim$0.6\% of the full-width at half maximum (FWHM).

\subsection{Theory}
Density Functional Theory (DFT) calculations presented in this work were performed using the Gaussian16 package.\cite{2016:gauss} The structures of perprotio and perdeutero phenylium were optimised at the B3LYP/N07D level,\cite{1993:becke,1988:lee,Barone:2008,Puzzarini2010} assuming singlet and triplet ground states and the double harmonic spectra were then calculated. The anharmonic quartic force field vibrational spectra were also calculated using generalised second-order vibrational perturbation theory at the B3LYP/N07D level.\citep{2012:BloinoBarone} This combination was shown to yield very good results for anharmonic frequency studies of PAHs.\cite{2018:Mackie} We used the keywords Opt=VeryTight and Int(Grid=200974) and default values for resonance thresholds. Further steps we took to improve the quartic force field anharmonic spectrum are described in detail in the ESI$^\dag$. In short, some low-energy normal modes do not exhibit the expected red shift when including anharmonic behaviour. Instead, these normal modes exhibited unusual blue shifts and were treated harmonically instead. This procedure resulted in a better agreement with experiments. All theoretical stick spectra were convoluted with a Gaussian line shape with a FWHM of 100 cm$^{-1}$ to match the experimental bandwidth. 

The PES describing the photodissociation of \ce{C6H5+} was investigated at the M06-2X/6-311++G(3df,2pd) level, considering both singlet closed-shell and triplet open-shell systems. For the latter, the HOMO and LUMO were mixed (Guess=Mix) to destroy $\alpha$-$\beta$ orbitals and spatial symmetries, and thus reproducing a correct UHF wave function. This specific level of theory was chosen because of the proven performances of the M06-2X functional coupled to a triple zeta quality basis set in predicting energy barriers and differences in energies for isomers in many systems,\cite{2008:zhao} including hydrocarbon rings and chains.\cite{2010:Steinmann} For example, the difference in energy between acetylene (C$_2$H$_2$) and vinylidene (CCH$_2$) is calculated to be 1.85~eV while the best theoretical prediction using coupled cluster methods is 1.86~eV.\cite{1996:Chang} Intermediates and transition states were found with the Berny algorithm and harmonic frequencies were calculated to verify the true nature of the stationary points (one imaginary frequency for transition states and zero for minima). Finally, Intrinsic Reaction Coordinate calculations were performed to check if transition states were connecting the correct minima. S-T gaps were evaluated for both \ce{C6H5+} and \ce{C6D5+} with electronic energies estimated at the M06-2X/6-311++G(3df,2pd) level and zero-point vibrational energies estimated at the anharmonic B3LYP/N07D level. The S-T gap for \ce{C6H5+-Ar} was estimated using M06-2X/6-311++G(3df,2pd) for both the electronic and zero-point vibrational energy, with the inclusion of empirical dispersion (GD3) which calculates the long-range potentials needed for the simulation of noble gas tagging.\cite{2010:grimme}

\section{Results and Discussion}

In the following, IRMPD spectra of both perprotio \ce{C6H5+} and perdeutero \ce{C6D5+} phenylium are presented, and compared to both harmonic and anharmonic DFT spectra to ascertain whether $^1$A$_1$ or $^3$B$_1$ is the phenylium ground state. IR fragmentation mass spectra have been recorded at different wavelengths, and are analyzed for both isotopologs to gain insight in the underlying loss mechanisms. Using high-level DFT calculations, the PES behind these loss mechanisms is explored starting from the experimentally confirmed ground state.  

\subsection{Spectroscopy}

\begin{figure*}
\includegraphics[width=\linewidth]{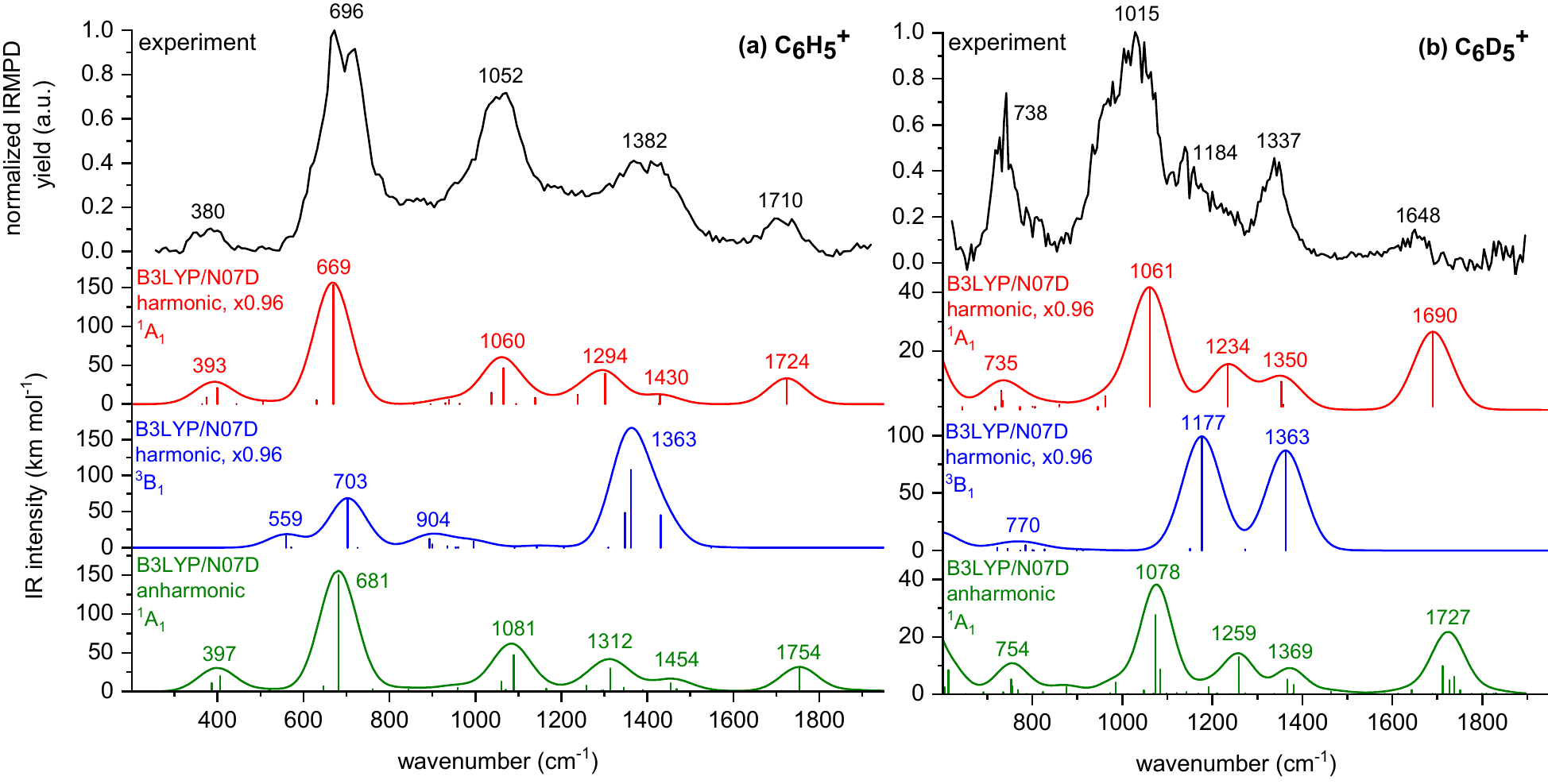}
\caption{ Comparison between the IR experimental spectrum of (a) phenylium-\ce{H5} and (b) phenylium-\ce{D5} (black curve, top), and the calculated singlet (red, second from top) and triplet (blue, third from top) spectra along with the anharmonic singlet (green, bottom) spectrum, for the respective isotopologs. The peak positions of the experimental spectrum were determined by applying Gaussian fits to the individual features, and using their maxima to describe the experiment. The harmonic curves were scaled with a factor of 0.96, while the anharmonic curve is unscaled. All theoretical spectra are convoluted with a 100 cm$^{-1}$ FWHM Gaussian waveform. \label{fig:IRphenyl}}
\end{figure*}

\begin{table*}
\small
 \caption{Observed band frequencies (cm$^{-1}$) and relative intensities together with the calculated singlet frequencies (cm$^{-1}$) and IR intensities (km$\cdot$mol$^{-1}$), and the descriptions of the corresponding vibrational modes for preprotio (left panel) and perdeutero (right panel) phenylium. The listed harmonic frequencies are scaled by a factor of 0.96. The abbreviations oop and ip stand for out-of-plane and in-plane.}
 \label{tbl:modes}
\begin{tabular*}{\linewidth}{@{\extracolsep{\fill}}lccccllcccc}
\cline{2-5} \cline{8-11}
            & \multicolumn{4}{c}{\ce{C6H5+}}                          &  &             & \multicolumn{4}{c}{\ce{C6D5+}}                          \\
            & \multicolumn{2}{c}{Harmonic calc.} & \multicolumn{2}{c}{Experiment} &  &             & \multicolumn{2}{c}{Harmonic calc.} & \multicolumn{2}{c}{Experiment} \\
Description & Freq.        & IR int.        & Freq.     & Norm. $Y$   &  & Description & Freq.        & IR int.        & Freq.     & Norm. $Y$   \\ \cline{1-5} \cline{7-11} 
\textcolor{RedOrange}{CCC$_{\text{oop}}$ empty C} & 375          & 9              & 380       & 0.10        &  & \textcolor{RedOrange}{CCC$_{\text{oop}}$ empty C} & 343          & 14             &           &             \\
\textcolor{Blue}{ring compression} & 399          & 21             &           &             &  & \textcolor{Blue}{ring compression} & 383          & 19             &           &             \\
ring twisting       & 505          & 3              &           &             &  &             &              &                &           &             \\
\textcolor{BurntOrange}{ring compression}& 630          & 5              &           &             &  & \textcolor{CadetBlue}{quintet CD$_{\text{oop}}$}       & 520          & 63             &           &             \\
\textcolor{CadetBlue}{quintet CH$_{\text{oop}}$}       & 699          & 153            & 696       & 1.00        &  & \textcolor{BurntOrange}{ring compression}& 588          & 7              &           &             \\
ring breathing      & 939          & 6              &           &             &  & CD$_{\text{ip}}$ scissoring     & 731          & 6              & 738       & 0.74        \\
ring breathing      & 1037         & 15             &           &             &  & ring compression       & 735          & 3              &           &             \\
CH$_{\text{ip}}$ scissoring     & 1065         & 46             & 1052      & 0.71        &  &             &              &                &           &             \\
CH$_{\text{ip}}$ scissoring     & 1139         & 8              &           &             &  & ring breathing      & 962          & 5              &           &             \\
CH$_{\text{ip}}$ rocking     & 1238         & 12             &           &             &  & CC stretch  & 1061         & 41             & 1015      & 1.00        \\
\textcolor{MidnightBlue}{CC stretch}  & 1302         & 39             & 1382      & 0.40        &  & \textcolor{MidnightBlue}{CC stretch}  & 1234         & 15             & 1184      & 0.30        \\
CC stretch  & 1430         & 12             &           &             &  & CC stretch  & 1353         & 10             & 1337      & 0.46        \\
\textcolor{Brown}{CC stretch}  & 1724         & 33             & 1710      & 0.14        &  & \textcolor{Brown}{CC stretch}  & 1690         & 27             & 1648      & 0.15        \\ \cline{1-5} \cline{7-11} 
\end{tabular*}
\end{table*}

The experimental IRMPD spectrum of perprotio phenylium is shown in Fig. \ref{fig:IRphenyl}a as the black solid curve (top). The spectrum is constructed using the four fragment masses shown in Fig. \ref{fig:ms}. It features five broad resonances between 300 and 1800 cm$^{-1}$, with a FWHM of roughly 80--100 cm$^{-1}$. These experimental features are listed in Table \ref{tbl:modes}. Depicted below the experimental curve are the predictions based on different types of calculations, all employing the B3LYP functional and N07D basis set. The red curve (second from the top) shows the harmonic spectrum for the singlet configuration, with the frequencies scaled by a factor of 0.96. 
The experimental spectrum agrees best with this singlet one, both in terms of overall shape and peak positions. A comparison is given in Table \ref{tbl:modes}, along with the vibrational mode descriptions. The equivalent triplet state (Fig. \ref{fig:IRphenyl} blue trace, third from the top) does not show any resemblance in overall shape. An indication of some contribution to the experimental spectrum might be inferred from the predicted 1363 cm$^{-1}$ resonance, which is very close to the observed 1382 cm$^{-1}$ feature. This feature is the least well reproduced for the singlet state. However, this could also be indicative of some poorly predicted modes. The limited experimental resolution, predictive power, and complex relationship between calculated linear absorption strengths and observed fragmentation, impede reliable quantification. A dominant singlet character can, however, clearly be inferred.

The anharmonic spectrum was also calculated using the same functional and basis set. The green, bottom curve in Fig. \ref{fig:IRphenyl}a shows the unscaled, anharmonic, spectrum calculated for the singlet state. It is quite similar to the scaled harmonic spectrum and provides a similarly reasonable match with experiment as the scaled harmonic spectrum. In comparison to the latter, it shows an improved match for the experimental 696 and 1382 cm$^{-1}$ features, while the deviation from the experimental 380 and 1710 cm$^{-1}$ features is increased. Overall, taking anharmonicity into account does not seem to lead to significant improvements compared to the harmonic spectrum. Most modes in the anharmonic spectrum appear blueshifted compared to the experimental spectrum. We attribute this to an experimental redshift caused by the high-intensity multiple-photon absorption.\cite{2003:oomens} The anharmonic calculation represents a linear absorption spectrum and will therefore not line up exactly with the high-intensity IR-MPD spectra reported here. Deviations between experiment and prediction can be masked more easily with the freedom to choose a scaling factor for the harmonic calculations.

We also present the IRMPD spectrum of perdeutero phenylium, \ce{C6D5+}, in Fig. \ref{fig:IRphenyl}b. The experimental spectrum (black, top) displays four broad bands between 600 and 1900 cm$^{-1}$. The 1015 cm$^{-1}$ band shows a large, high-frequency shoulder. A deconvolution into two Gaussian curves yields band maxima at 1015 and 1184 cm$^{-1}$. Best agreement is again found between the experimental spectrum and the harmonic prediction for the singlet configuration (red, second from the top). The dominant feature in the experimental spectrum -- at 1015 cm$^{-1}$ -- is mirrored by the intense band at 1061 cm$^{-1}$, whereas no activity is predicted at these frequencies for the triplet state (blue, third from the top). The 738 cm$^{-1}$ and 1648 cm$^{-1}$ experimental features are quite accurately predicted by the singlet spectrum, where the triplet spectrum shows only little or no intensity. Overall, the shape and band positions agree best with the singlet ground state prediction. Although again a triplet contribution to the spectrum cannot be excluded, the phenylium in its $^1$A$_1$ state is clearly the dominant species observed, confirming our interpretation of the perprotio spectrum. Just as for the perprotio phenylium, the unscaled, anharmonic singlet spectrum (green, bottom curve) shows great similarity with the scaled harmonic prediction, but does not yield a much improved fit for any of the observed features.

The experimental features and the scaled, harmonic resonances are presented in Table \ref{tbl:modes}. Comparison between the perdeutero and perprotio phenylium reveals an expected red-shift of the vibrational modes. Where the perprotio phenylium exhibits modes with mixed CH${_{\text{ip}}}$ bending and CC stretching character in the 1000--1600 cm$^{-1}$range, in perdeuterated phenylium, the CD$_{\text{ip}}$ counterpart is less prominent than the CC stretching component. We compared the nature of these modes between \ce{C6H5+} and \ce{C6D5+} by inspecting their animated vibrations. Many of these mixed modes in the \ce{C6H5+} spectrum have no one-on-one equivalent counterpart with significant intensity in the \ce{C6D5+} spectrum. Of the modes presented in the Table, six show direct counterparts with significant intensity. They are indicated with matching colors.

Our calculations of the S-T gap for perprotio phenylium with DFT calculations (using M06-2X for the electronic energies and B3LYP for the zero-point energies) yield a value of 121 kJ mol$^{-1}$ (120 kJ mol$^{-1}$ for \ce{C6D5+}), close to the highest values reported in literature ($\leq 137$ kJ mol$^{-1}$\cite{2016:Peverati}). The S-T gap of argon-tagged phenylium \ce{Ar - C6H5+}, calculated  including empirical dispersion to account for the long-range potentials needed to describe argon-tagging,  is 99 kJ mol$^{-1}$, whereas a dispersion-corrected calculation for bare phenylium yielded 91 kJ mol$^{-1}$. This indicates a small stabilization of 8 kJ mol$^{-1}$ for the singlet state by the $\sigma$-coordinated --- or in-plane --- argon atom, as was previously suggested as well.\cite{2010:patzer}

We conclude that the spectra for both perprotio and perdeutero systems clearly indicate that the observed phenylium is in its  $^1$A$_1$ singlet ground state. DFT calculations confirm these findings.

\subsection{Fragmentation chemistry}

\begin{figure}
\centering
\includegraphics[width=0.9\linewidth]{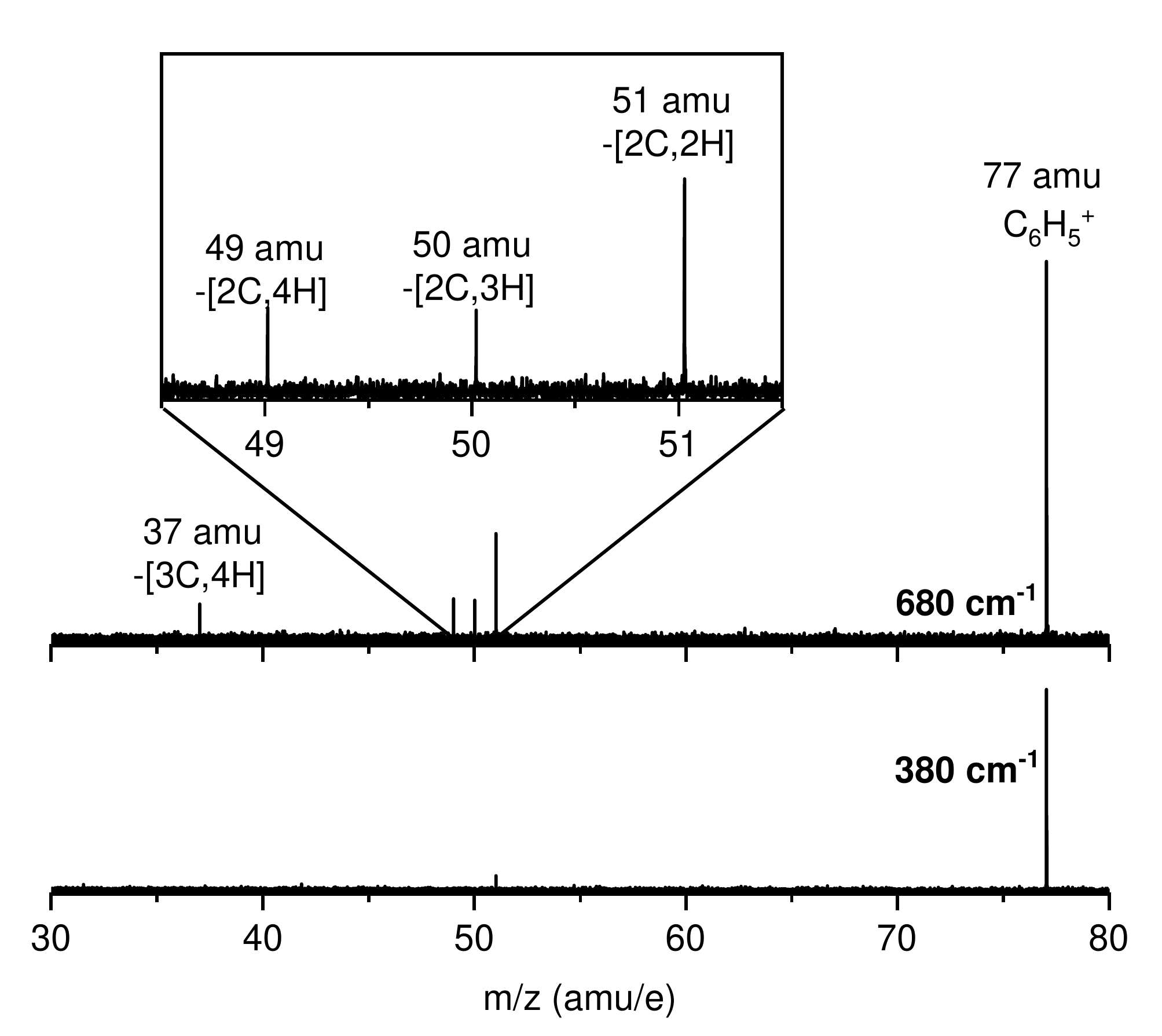}
\caption{IRMPD mass spectra of perprotio phenylium (\ce{C6H5+}, $m/z=$77), recorded at 680 cm$^{-1}$ (top) and 380 cm$^{-1}$ (bottom). \label{fig:ms}}
\end{figure}

\begin{figure}
\centering
\includegraphics[width=0.9\linewidth]{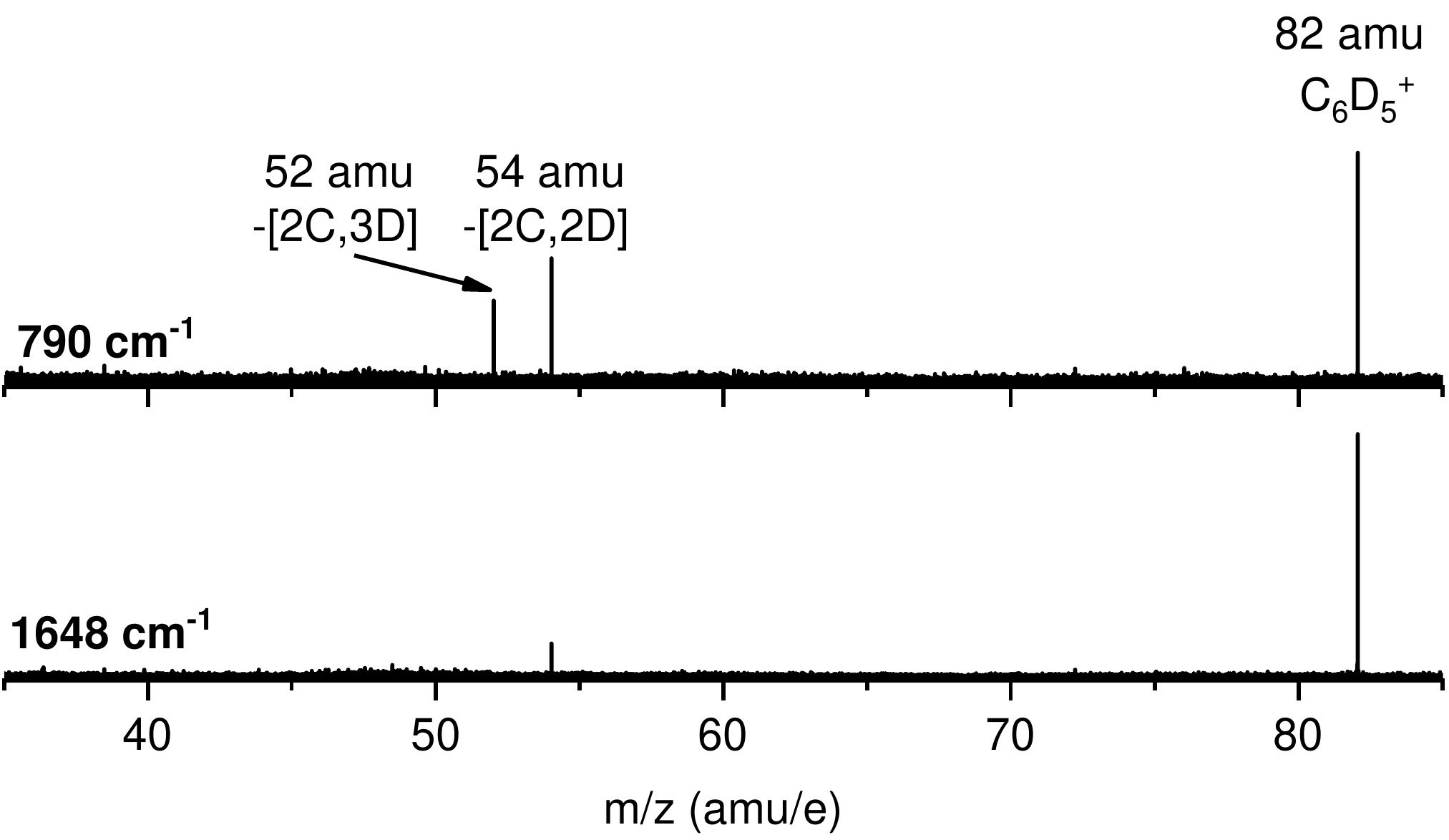}
\caption{IRMPD mass spectrum of perdeutero phenylium (\ce{C6D5+}, $m/z=$82), recorded at 790 cm$^{-1}$ (top) and 1648 cm$^{-1}$ (bottom). \label{fig:msd5}}
\end{figure}
 
Upon resonant IR irradiation, several fragment masses appear. Figure \ref{fig:ms} shows mass spectra recorded at two different resonances in the IR spectrum of perprotio phenylium. The top panel shows the fragmentation that occurred close to the maximum of the 696 cm$^{-1}$ band at 680 cm$^{-1}$. Four fragment ions are observed at $m/z=$51, 50, 49 and 37 in addition to the parent ion at $m/z=77$. The mass spectrum recorded at the weakest IR resonance at 380 cm$^{-1}$ (bottom panel) displays only one fragment at $m/z=51$. Fragmentation mainly occurs through the loss of [2C,nH]-moieties with $n=2-4$ ($m/z=51-49$) plus a small contribution of [3C,4H]-loss ($m/z=37$). The $m/z=50$ and 49 channels appear mostly along this dominant $m/z=51$ mass, while the $m/z=37$ channel only appears at the most intense resonances. Notably, no significant H or 2H loss ($m/z = 76$ and $75$) is observed for any of the resonances. This was also observed in an earlier IRMPD measurements of naphthyl cation, showing [2C,2H] as the dominant loss channel\cite{2012:alvarogalue} although the relatively low mass resolution in that experiment could have obscured observation of H-loss.

For perdeutero phenylium, a mass spectrum recorded at 1275 cm$^{-1}$ --- on the flank of the IR resonance at 1337 cm$^{-1}$ -- - is given in the top panel of Fig. \ref{fig:msd5}, displaying two main fragment masses at $m/z=54$ and 52 next to the parent mass at $m/z=82$. Contributions from \ce{[nD]}-loss at $m/z=80$, 78 and 76 are small and are only present in the flanks of the 1337 cm$^{-1}$ and 1015 cm$^{-1}$ resonances. No $m/z=38$ ion signal is detected, which would have indicated [3C,4D]-loss. The mass spectrum recorded at the weakest IR resonance (1648 cm$^{-1}$) is displayed in the bottom panel and only shows one fragment mass, at $m/z=54$. Similar to its perprotio counterpart, perdeutero phenylium mainly loses [2C,2D] ($m/z=54$), with [2C,3D] ($m/z=52$) appearing along the main fragment in all resonances other than the weakest 1648 cm$^{-1}$ feature. 

Both the perprotio and perdeutero phenylium thus show the loss of [2C,2H/D] as their dominant loss channel, with the loss of [2C,3H/D] as the second strongest. The most salient differences are that the perprotio phenylium is the only one to show the loss of [3C,4H/D], while the perdeutero phenylium is the only one to show the loss of \ce{[nH/D]}. In spite of these differences, it is clear that these smaller channels are secondary to the loss of [2C,2H/D]. We attribute the observation of lower-mass fragments for \ce{C6H5+} to a higher-energy loss channel, that is made accessible by a combined effect of both the IR absorption strength of the excited mode and laser intensity.

\begin{figure*}
\includegraphics[width=0.9\linewidth]{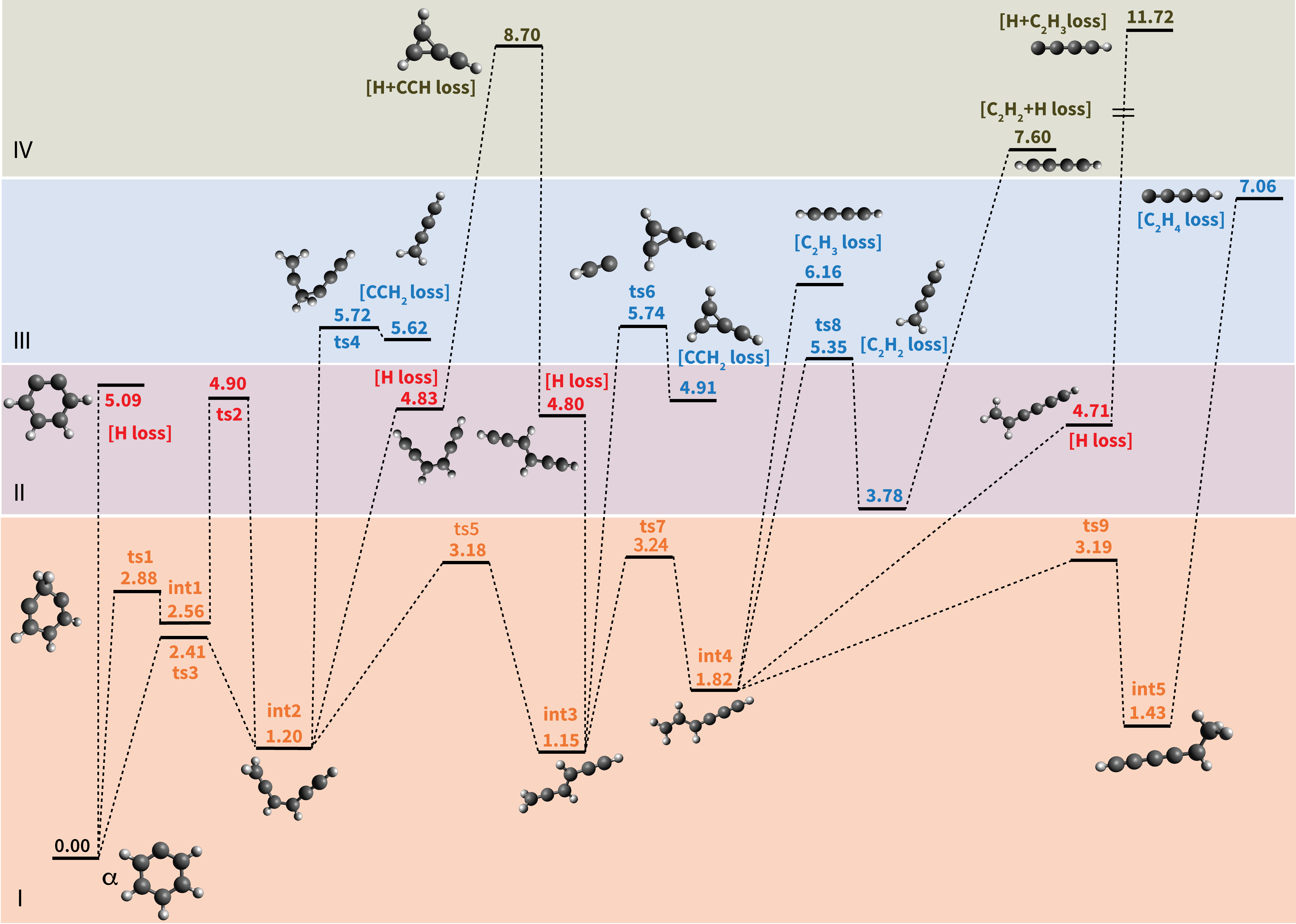}
\caption{Potential Energy Surface (PES) connecting phenylium and several low-lying isomers with potential fragmentation pathways (M06-2X/6-311++G(3df,2pd) level of theory). Energies are in eV and calculated with respect to phenylium, loss channels are labeled between parentheses. Molecular structures depicted in the PES are cations. The colors (background and labels) and roman numerals identify energy regimes associated with different processes going from low to higher energy: isomerization  (0 to $\approx$~4~eV, I),  H-loss (4 to $\approx$~5~eV, II), C-loss (5 to $\approx$ 7~eV III) and sequential C- and H-loss (above 7~eV, IV).  The PES of \ce{C6D5+} will be similar with minor differences in the energetics due to zero point vibrational energy differences between perprotio and perdeuterated phenylium. \label{fig:PES}}
\end{figure*}

To rationalize the observed mass spectra obtained upon IR irradiation,  we investigated the PES describing the photofragmentation of phenylium at the M06-2X/6-311++G(3df, 2pd) level. The results are summarized in Fig.~\ref{fig:PES} where the colors and roman numerals highlight energy regimes dominated by a given mechanism. The lowest energy regime is that of ring opening and subsequent isomerization processes. Ring opening (I, orange) is clearly energetically more favorable than H-atom loss (II, red) processes. Direct H-loss from phenylium involves a simple \ce{C-H} bond breaking requiring 5.09~eV. This value is slightly higher (0.2~eV) than obtained at the CCSD(T) level,\cite{2016:Peverati} which is likely due to the multi-configuration character of the {\it m}-benzyne \ce{C6H4+} cation.\cite{2008:Li}
Ring opening --- leading to the chain-like intermediate {\bf int2} --- can occur along two different reaction paths. The lowest energy path involves a single transition state {\bf ts3} at 2.41~eV, where the breaking of the alpha \ce{C-C} bond is followed by the migration of one H-atom to form a CH$_2$-moiety. The higher energy path goes through {\bf ts1} at 2.88~eV involving the migration of an H atom to a nearby carbon creating {\bf int1}. This is followed by the subsequent cleavage of the \ce{C-C} bond with a concerted migration of one H to the final carbon of the chain, involving {\bf ts2} at 4.9~eV. In both paths the transition states and the intermediate have a singlet open-shell character. The 2.41~eV energy of {\bf ts3} compares very well with the value of 2.44~eV (56.2 kcal/mol) obtained with CCSD(T)\cite{2016:Peverati} while it is low compared to the 3.33~eV value obtained with QCISD(T).\cite{Shi:2016} Despite the uncertainties of the different methods, direct H-loss from phenylium is clearly energetically less favorable than isomerization.

Several chain-like \ce{C6H5+} isomers exist, lying a couple of eVs above ground-state phenylium.\cite{Shi:2016} The structures investigated in Fig.~\ref{fig:PES} are amongst the lowest in energy. Similar reaction paths were found to be available for the all isomers investigated here, namely {\bf int2}, {\bf int3}, {\bf int4}, and {\bf int5}, with comparable energies. In the following, we describe the different reaction pathways available for isomer {\bf int3} as an example, the energies are given with respect to {\bf int3}. For this isomer, the lowest energy pathways are isomerization reactions (I, orange): a rotation along the \ce{C2H2} central group ({\bf ts5}, 2.03~eV) producing the {\it cis} form of {\bf int3} --- {\bf int2}, and H-migration ({\bf ts7}, 2.09~eV) leading to {\bf int4}. An additional H-transfer leads to the formation of a methyl group in {\bf int5}.

Energetically, the next class of reactions is H-loss (II, red), requiring energies ranging from 3.65~eV to 4.8~eV depending on which H-atom is eliminated.\footnote{In Fig.\ref{fig:PES} we reported only the lowest energy H-loss channel for each intermediate.}
The \ce{CCH2}-loss reaction (III, blue) is found to be even more endoergic, requiring energies above 5.74~eV ({\bf ts6}) for the direct dissociation and above 8.70~eV for the sequential H+CCH loss (IV, green). Direct CCH$_n$ losses (with $n=3,4$) also require energies higher than 6~eV.

Based on the PES, the following scenario can be put forward. When phenylium is excited up to an energy of around 4~eV, isomerization reactions (such as ring opening/closing, H-migration, and rotation) can occur. Because the isomerization barriers are energetically very close  and lower than the dissociation barriers the process is expected to yield a mixed population of \ce{C6H5+} isomers, including phenylium.

\begin{figure}
\centering
\includegraphics[scale=0.39]{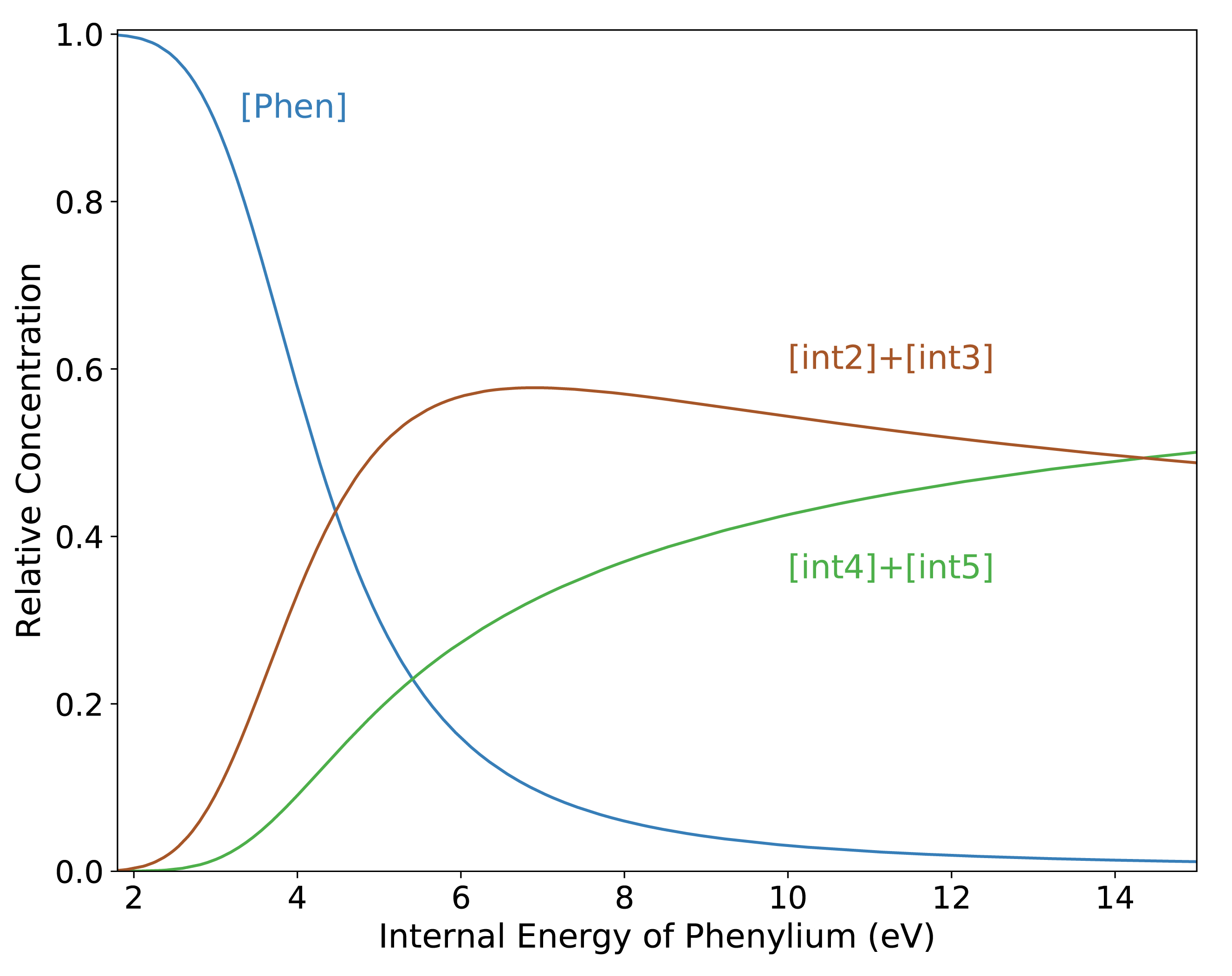}
\caption{Relative concentrations of phenylium, {\bf int 2}+{\bf int 3} and {\bf int 4}+{\bf int 5} in the equilibrated mixture M as function of the internal energy of the phenylium molecules in the pre-equilibrium approximation. \label{fig:isomers}}
\end{figure}

This situation can be described quantitatively within the pre-equilibrium approximation.\cite{pre-equi} In this approximation, the relative abundance of each isomer in an equilibrated mixture of isomers can be determined via the vibrational density of states (VDOS) at internal energy E.\cite{2015:Solano} The abundance of isomer X can be calculated as \[ [X] = \rho_X(E-E_{0,X}) / \sum_{i} \rho_i(E-E_{0,i})\]where $\rho_{i}(E-E_{0,i})$ is the density of states of isomer i evaluated at energy E, with E$_{0,i}$, the relative energies of isomer i, all given with respect to the ground state of phenylium as a reference. For each isomer, the VDOS was calculated using the direct count method on unscaled harmonic frequencies from M06-2X/6-311+G(3df,2pd) calculations. The resulting abundances are plotted in Fig. \ref{fig:isomers}. Phenylium constitutes 90\% of this mixture up to 3~eV of internal energy, but already around 4~eV its contribution is down to 50\%. From this point, the sum of int~2 and int~3 dominates the mixture of isomers up to 14~eV, where their contribution decreases below 50\% and isomers of int~4 and int~5 dominates.
Thus it is a mixture of \ce{C6H5+} isomers that will determine the fragmentation observed experimentally,  assuming that at intermediate internal energies of several eVs all species are in the quasi-continuum$^{\dag}$\cite{Simpson1985} and readily absorb the extra photons needed for fragmentation. It appears that the ring opening in phenylium brings a thermodynamic advantage: int~2 and int~3 possess more low-energy vibrational modes that increase their density of states with respect to phenylium. A similar behavior was also observed for the naphthalene cation.\cite{2015:Solano}

Starting at an energy of around 4.8~eV, H-loss reactions from a mixed population of \ce{C6H5+} isomers become accessible. Direct H-loss from the ring might also happen, but it is not expected to be competitive with H-loss from the population of open isomers. Finally CCH$_2$/C$_2$H$_2$-loss reactions open up for energies above 5.8~eV.

This theoretical picture appears to be at odds with what was observed in the experiment, which reveals only C-loss channels.
Two studies on the reactivity of phenylium showed only the loss of [2C,2H] as a collisionally induced dissociation (CID) byproduct at room temperature.\cite{Ascenzi2007,Giordana2009} In these guided ion beam studies, the collisions are expected to lead to a less gentle excitation than in our IR-MPD experiment, but the lack of hydrogen loss appears consistent with the current observations.
In contrast, for the neutral phenyl radical, photofragmentation studies using 248 nm light (5.0 eV per photon)  showed H-loss only, whereas additional [2C,2H]-loss was observed when using 193 nm light (6.4 eV).\cite{Negru2010} Although these findings cannot be directly compared due to the difference in charge state, it is interesting to note that the shift from H-loss to [2C,2H]-loss occurs in the range between 5.0 and 6.4~eV, the same range calculated here for the phenyl ion.

The lack of detection of the H-loss channel may be explained if the [2C,nH]-peaks in the mass spectra are the result of sequential loss of H followed by the loss of [2C,$(\text{n}-1)$H]-units. In this scenario, isomerization takes place first producing a collection of isomers, within the microsecond timescale of the experiment. In any of these isomers, fast loss of [2C,$(\text{n}-1)$H] follows the initial loss of H-atoms,  thus obscuring this channel. 

The high barriers calculated for the [2C,H]-loss reactions from \ce{C6H4+} isomers plus the sole detection of [2C,2H] in the CID experiments might suggest that another mechanism is at play. It could also be speculated that a kinetic shift might be responsible for the lack of detection of H-loss on \ce{C6H5+}, {\it i.e.}, that a molecule needs to be excited to higher internal energies to observe any fragmentation during the experimental timescale ($\sim\mu$s). This has been observed both in large and small aromatic systems.\cite{2010:bakker,West2018} At these higher internal energies ($\geq6$~eV) the dominant fragmentation channel might be different from that at lower internal energies.

To evaluate if there is a change in the dominant fragmentation channel at higher energies we calculated entropies of activation at T=1000~K  ($\Delta$S$_{1000}$) for H-loss and \ce{CCH2}-loss from {\bf int2}, using the unscaled harmonic frequencies for {\bf int2} and {\bf ts4}.\cite{1997:Baer} Since H-loss is a barrierless reaction, we started from an optimised structure where the eliminated H is located 4~\AA~away from the remaining \ce{C6H4+} fragment.
For the H-loss reaction, $\Delta$S$_{1000}$ is 23.02 J~mol$^{-1}$K$^{-1}$, while for \ce{CCH2}-loss $\Delta$S$_{1000}$ is 51.58 J~mol$^{-1}$K$^{-1}$. This means that, even if the loss of an H atom requires a lower energy barrier (3.63~eV) than the loss of \ce{CCH2} (4.52~eV), the reaction rate of \ce{CCH2}-loss will grow faster than the rate of H-loss as function of internal energies due to the larger change in entropy. Eventually the \ce{CCH2}-loss channel becomes dominant at high internal energy. A similar situation was observed for the competition between H- and \ce{H2}-loss channels in the pyrene cation.\cite{1995:Ling}
This is true also for \ce{C6D5+}, since the calculated $\Delta$S$_{1000}$ for D-loss and \ce{CCD2}-loss are 27.21 and 55.65 J~mol$^{-1}$K$^{-1}$, respectively. 

The lack of detection of the H-loss channel in \ce{C6H5+} could thus be either due to sequential loss or to a kinetic shift. Experiments on a nanosecond timescale could disentangle the direct and/or sequential origins of the C-loss channels. 
A kinetic modeling study of the fragmentation pathways of phenylium and its isomers, which is beyond the scope of this paper, could help shed light on this matter.

\section{Conclusions}
The first IR spectra of isolated, gas-phase perprotio phenylium and perdeutero phenylium covering the 5.2--38 $\mu$m range are presented. Our experiments confirm the assignment of the singlet ground-state configuration and reveal only limited anharmonic behavior. The \ce{C6H5+} ions fragment primarily through loss of [2C,nH] ($n=2-4$), with [2C,2H]-loss being the dominant loss channel, in agreement with other works. The exploration of the PES of phenylium through DFT calculations and pre-equilibrium calculations of the VDOS for various intermediates revealed facile ring-opening pathways, suggesting a mixture of open isomers to be available upon photolysis.  H-loss was not detected in experiment. This could either be explained by fast sequential loss or by kinetic effects; the entropy  ($\Delta$S$_{1000}$) calculations show that at high internal energies, \ce{CCH2}-loss proceeds at a faster rate than H-loss. 
The presented results constitute the successful IRMPD spectroscopy of an IRMPD-resistant ion using the FELICE FT-ICR MS. Developments are currently in progress that improve ion production, storage, and alignment, providing access to higher ion and photon densities. The current results demonstrate the unique capabilities of FELICE in spectroscopic characterization of strongly bound species which would not be possible with other instruments, thus paving the way to the IRMPD spectroscopic characterization of highly photostable, astronomically relevant species down to the far-infrared.

\section*{Author contributions}
A.P. and J.O. conceptualized the study.  A.P. and A.G.G.M.T. acquired the funding for the work. S.D.W, J.R.E., A.P., J.M.B. and G.B performed the experiments. A.C. performed the theoretical calculations. S.D.W, A.P. and A.C. conducted the data analysis. S.D.W., A.C. and A.P. wrote the manuscript. All authors contributed to the scientific discussion and gave critical reviews of the manuscript.

\section*{Conflicts of interest}
There are no conflicts to declare.

\section*{Acknowledgements}
We gratefully acknowledge the {\it Nederlandse Organisatie voor Wetenschappelijk Onderzoek} (NWO) for the support of the FELIX Laboratory. In particular, we would like to thank A. F. G. van der Meer and B. Redlich of the FELIX Laboratory for their support during the development of the FELICE FT-ICR apparatus.  We would also like to thank Prof. W. J. Buma and two anonymous reviewers for their constructive comments. This work is supported by the VIDI grant (723.014.007) of A.P. from NWO. Furthermore, A.C. gratefully acknowledges NWO for a VENI grant (639.041.543). Calculations were carried out on the Dutch national e-infrastructure (Cartesius and LISA) with the support of Surfsara, under projects NWO Rekentijd 16260 and 17603.

\bibliography{rsc} 
\bibliographystyle{rsc} 

\end{document}